\documentclass[twocolumn,pra,showpacs]{revtex4}

\usepackage{graphicx}
\usepackage{rotating}
\usepackage{amsmath}
\usepackage{amsfonts}
\usepackage{amssymb}
\usepackage{enumerate}
\usepackage{longtable}
\usepackage{dcolumn}
\usepackage{bm}

\begin{document}

\title{Thermodynamics of Adiabatically Loaded Cold Bosons in the
  Mott Insulating Phase of One-Dimensional Optical Lattices}

\author{K.P. Schmidt$^1$}
\email{kaiphillip.schmidt@epfl.ch}
\author{A. Reischl$^2$}
\author{G.S. Uhrig$^3$}
\affiliation{$^1$ Institute of Theoretical Physics, \'{E}cole Polytechnique F\'{e}d\'{e}rale de Lausanne, CH 1015 Lausanne, Switzerland}
\affiliation{$^2$ Institut f\"ur Theoretische Physik, Universit\"at zu K\"oln,
 Z\"ulpicher Stra{\ss}e 77, 50937 K\"oln, Germany}
\affiliation{$^3$ Theoretische Physik, FR 7.1, Geb.\ 38, 
Universit\"at des Saarlandes, 66123 Saarbr\"ucken, Germany}
\date{\rm\today}

\begin{abstract} 
In this work we give a consistent picture of the thermodynamic properties of
bosons in the Mott insulating phase when loaded adiabatically into
one-dimensional optical lattices. We find a crucial dependence of the
temperature in the optical lattice on the doping level of the Mott
insulator. In the undoped case, the temperature is of the order of the large 
onsite
Hubbard interaction. In contrast, at a finite doping level the
temperature jumps almost immediately to the order of the small hopping
parameter. These two situations are investigated on the one hand by considering
limiting cases like the atomic limit and the case of free fermions. On the
other hand, they are examined using a quasi-particle conserving continuous 
unitary  transformation extended by an approximate thermodynamics  for
 hardcore particles.       
\end{abstract}



\maketitle

\section{Introduction}
The  design of tunable quantum systems with strong correlations has
attracted much interest due to the fast experimental developments in the field
of ultracold atoms loaded in optical lattices. It has become possible in
experiment to vary the dimension of the optical lattices and the particle
statistics of the involved ultracold atoms
\cite{jaksc98,grein02,stofe04,pared04,kinos04,kohl05a}. 
This opens the road to study fascinating quantum many-body effects like 
superfluidity in artificially created and thus tunable structures.

In order to investigate these quantum effects the temperature of the ultracold
atoms in the optical lattice has to be small in units of typical model
parameters. Experimentally it is an open problem to measure this temperature. 
We have addressed this question in a recent work \cite{reisc05} theoretically 
by analyzing spectroscopic experiments on bosons in one-dimensional optical 
lattices \cite{stofe04}. We  found that the temperature in the Mott insulating 
phase is surprisingly large,  namely of the order of the onsite Hubbard 
interaction $U$. In an analogous  experiment the so-called Tonks-Girardeau 
regime (TG) was  investigated \cite{pared04}. From this experiment, a rather 
small temperature of the order of  the hopping parameter $t$ has been deduced.

In the present work we provide a consistent thermodynamic picture of both 
experiments. The large difference in temperature can be understood solely
from entropy considerations. In the undoped Mott insulator the groundstate is
unique and does not carry any entropy. It is separated from excited states by
an energy gap which is determined by the interaction $U$. The entropy 
increases appreciably only at temperatures which are of the order of the gap. 
In a \emph{doped} Mott insulator  describing the TG regime the low-lying 
states produce a finite entropy and the temperature of the bosons is much 
smaller.

The paper is organized as follows. We will first introduce the model and our
notation. In Sect.~\ref{sec:limits} we will study several analytically
solvable limits. First, we calculate the entropy of non-interacting bosons in
a trap relevant for the cold atoms before loading into the optical lattice. 
Then we consider the atomic limit and the limit of non-interacting fermions 
which corresponds to the TG case. In Sect.\ \ref{sec:cut} a continuous unitary 
transformation (CUT) is used to study the
Mott insulating phase of the Bose-Hubbard model at zero temperature. The
results are extended to finite temperatures by an approximate formalism 
appropriate for hardcore particles. We first test the reliability of this 
approximation by comparison to exact results in the case of non-interacting 
fermions. Then the thermodynamic properties of the doped and undoped Mott 
insulating phase of the one-dimensional Bose-Hubbard model are studied. The 
conclusions will summarize the essence of the paper. In the last section,
we propose an experimental procedure to reach very low temperatures also
for the Mott insulating phase.
          
\section{Model}
\label{sec:model}

Bosonic atoms in an optical lattice are described by the Bose-Hubbard model
\cite{jaksc98}. The Hamiltonian reads
\begin{eqnarray}
\label{eq:bose:hamiltonian}
H &=& t H_{t}+ U H_{U} + \mu H_{\mu} \\\nonumber
  &=& -t\sum\limits_i( b^\dagger_i b^{\phantom{\dagger}}_{i+1}+b^\dagger_{i+1}
       b^{\phantom{\dagger}}_i )\\\nonumber
  &+& (U/2)\sum_i \hat{n}_i ( \hat{n}_i-1) 
       -\mu \sum_i \hat{n}_i
\end{eqnarray}
where the first term is the kinetic part $t H_{t}$ and the second term the
repulsive interaction $U H_{U}$ with $U>0$. The last term accounts for a
chemical potential $\mu$. 
The bosonic annihilation (creation)  operators are denoted by 
$b^{(\dagger)}_i$, the number of bosons by
$\hat{n}_i=b^\dagger_i b^{\phantom{\dagger}}_i$. 
The average number of bosons is $n=\langle \hat{n}_i\rangle = N/L$ where
$N$ is the total number of bosons and $L$ the total number of wells in the
optical lattice. In the spirit of the tight-binding description we will call
the wells henceforth `sites'.
On the same site, the bosons
repel each other with interaction strength $U$.

In the limit $t/U \to 0$, a site with $n$ bosons contributes the energy 
$E(n)=(U/2) n(n-1)$. A deviation from uniform filling costs the energy
$E(n+1)+E(n-1)-2E(n)=U$. Note that this excess energy results from the constant
curvature of the parabola $\propto n^2$. Thus it is independent of $n$.
For large $U$ and integer filling $n$ it is  energetically favorable to 
distribute the bosons uniformly and to put just $n$  bosons on each site. The 
system is in an insulating state with an energy gap of the order of $U$.  

In the opposite limit of large $t/U$ the bosons form a superfluid condensate 
for low enough temperatures. 

We will not address the inhomogeneity that is present in the
experimentally realized traps, i.e., we treat the system as translationally
invariant though there is the slowly varying external trap potential.
This approach makes quantitative computations much easier and it is backed by
the recent finding that the inhomogeneous system acts essentially like
two (or several) independent systems of different average number of bosons
\cite{batro05}.

Among the  extensive literature on the bosonic Hubbard model there are
mean-field treatments, e.g.~Refs.~\onlinecite{jaksc98,kraut92,shesh93,ooste01},
other analytical investigations, e.g.\ Refs.\
\onlinecite{fishe89,elstn99,kraut91,dicke03}, and numerical treatments, e.g.\
Refs.\ \onlinecite{pai96,kashu96,kuhne98,kuhne00,reisc05}.
The phase diagram in the $t$-$\mu$-plane consists of a series of lobes. 
The lobes of the Mott insulating phases correspond
to an integer filling per site. The present paper is only concerned
with the first lobe of the phase diagram where $n=1$.

\section{Limiting Cases}
\label{sec:limits}

Three limiting cases are investigated in order to understand
the basic entropy properties of the cold atoms before and after adiabatic
loading into the optical lattice. First, we calculate the entropy of
non-interacting bosons in a trap. This entropy is present before turning on the
optical lattice. It can be viewed as a function of the superfluid fraction. 
The same entropy has to be realized in the optical lattice once it has been
turned on in an adiabatic fashion. 
Hence the adiabaticity of the loading determines the 
temperature of the bosons in the optical lattice \cite{blaki04}.  
We calculate the entropy as a function of the microscopic 
model parameters and of the doping in the atomic limit, 
$t/U=0$ and in the limit of infinite $U$, the so-called TG regime. In one 
dimension \cite{jorda28,tonks36,girar60},
 a Jordan-Wigner mapping to non-interacting fermions is appropriate.

\subsection{Free Bosons}
\label{ssec:bosons}

The formula for the entropy of free bosons in a wide
harmonic trap is derived. The Hamiltonian of the harmonic oscillator in $d$ 
dimensions reads ($\hbar=1$)
\begin{equation}
  \label{eq:app:en:HO}
  H= \omega_0 \sum_{j=1}^d a^\dagger_j a^{\phantom{\dagger}}_j.
\end{equation}
Many bosons loaded in the harmonic trap are considered. The energy $\omega_0$ 
is small. Therefore, sums over multiples of $\omega_0$ can be rewritten as 
integrals. The groundstate is macroscopically occupied with a fraction
\begin{equation}
  \label{eq:app:en:f0}
  f_0 = N_0/N
\end{equation}
where $N$ is the total number of bosons and $N_0$  their number
in the condensate. With the density of states $D(\omega)$ and the boson 
occupation $n_\text{B}(\omega)=1/(e^{\beta \omega}-1)$
one has
\begin{subequations}
\begin{eqnarray}
  N&=& N_0 + \sum_{m=1}^\infty n_\text{B}(m\omega_0) D(m\omega_0)
\\
\label{eq:app:en:N}
&=&N f_0 + \int_0^\infty n_\text{B}(\omega) D(\omega) d\omega \ .
\end{eqnarray}
\end{subequations}
The density of states $D(\omega)$ counts the number of states in a given energy
interval. Let us consider intervals of length $\omega_0$ centered around the 
energies $m \omega_0$, where $m$ is a non-negative integer.
We consider the number of ways in which the energy 
$\omega = m \omega_0$ can be realized. One has to distribute $m$ 
indistinguishable energy
quanta over $d$ oscillators. The number of possibilities is 
\begin{equation}
  \label{eq:app:en:countD}
  D(m\omega_0)\omega_0 = \binom{m+d-1}{d-1}\ .
\end{equation}
Using $m=\omega/ \omega_0$, the leading power in $\omega$ reads
\begin{equation}
  \label{eq:app:en:D}
  D(\omega)=\frac{1}{\omega_0(d-1)!} 
\left( \frac{\omega}{\omega_0}\right)^{d-1}\ .
\end{equation}
Only the leading order in $m=\omega/\omega_0$ is retained because a 
\emph{wide} trap potential is considered which implies a small oscillator
frequency $\omega_0$.

Specializing to our case of interest, a three-dimensional trap ($d=3$),
Eq.\ \ref{eq:app:en:N} becomes
\begin{equation}
  \label{eq:app:en:Nint}
  N=N f_0 + \frac{T^3}{2}\int_0^\infty \frac{x^2}{e^x -1} dx=N f_0 + T^3 
\zeta(3),
\end{equation}
where $x=\beta \omega$ and $\zeta(z)$ is the Riemann $\zeta$-function.
Furthermore, we need the total entropy. The entropy of a single bosonic
level $\epsilon$ reads ($k_{\rm B} =1$)
\begin{equation}
  \label{eq:app:en:S2}
  S(\epsilon)=-\ln(1-e^{-\beta\epsilon})+\beta\epsilon n_\text{B}(\epsilon)\ .
\end{equation}
The total entropy is given by
\begin{equation}
  \label{eq:app:en:Stot}
  S^\text{tot}= \int_0^\infty S(\epsilon)D(\epsilon)d\epsilon =
\frac{2\pi^4}{45}\left(\frac{T}{\omega_0}\right)^3\ .
\end{equation}
The condensate at zero energy does not contribute to the entropy 
because it represents a single, pure quantum state. Combining
Eqs.\ (\ref{eq:app:en:Nint}) and (\ref{eq:app:en:Stot}) leads to
\begin{equation}
  \label{eq:app:en:SproN}
  S/N=(1-f_0) \frac{2\pi^4}{45\zeta(3)}\approx 3.60 (1-f_0)\quad .
\end{equation}
This relation between entropy and superfluid fraction will serve in the
following as the initial entropy of the cold atoms before turning on 
the optical  lattice.

\subsection{Atomic Limit}
\label{ssec:atomic}

In this part, we calculate the entropy as a function of temperature and doping 
in the atomic limit, namely $t/U=0$. Then, all excitation processes are 
completely local. For simplicity, we restrict the local Hilbert space to 
the state with occupation one and the two energetically adjacent states with 
occupation zero and two. The energy to create a site with two bosons is $U$. 
All states with an even larger number of bosons have a higher energy and are 
therefore not important for the entropy in the temperature range considered 
here.

\begin{figure}[t]
  \begin{center}
    \includegraphics[width=\columnwidth]{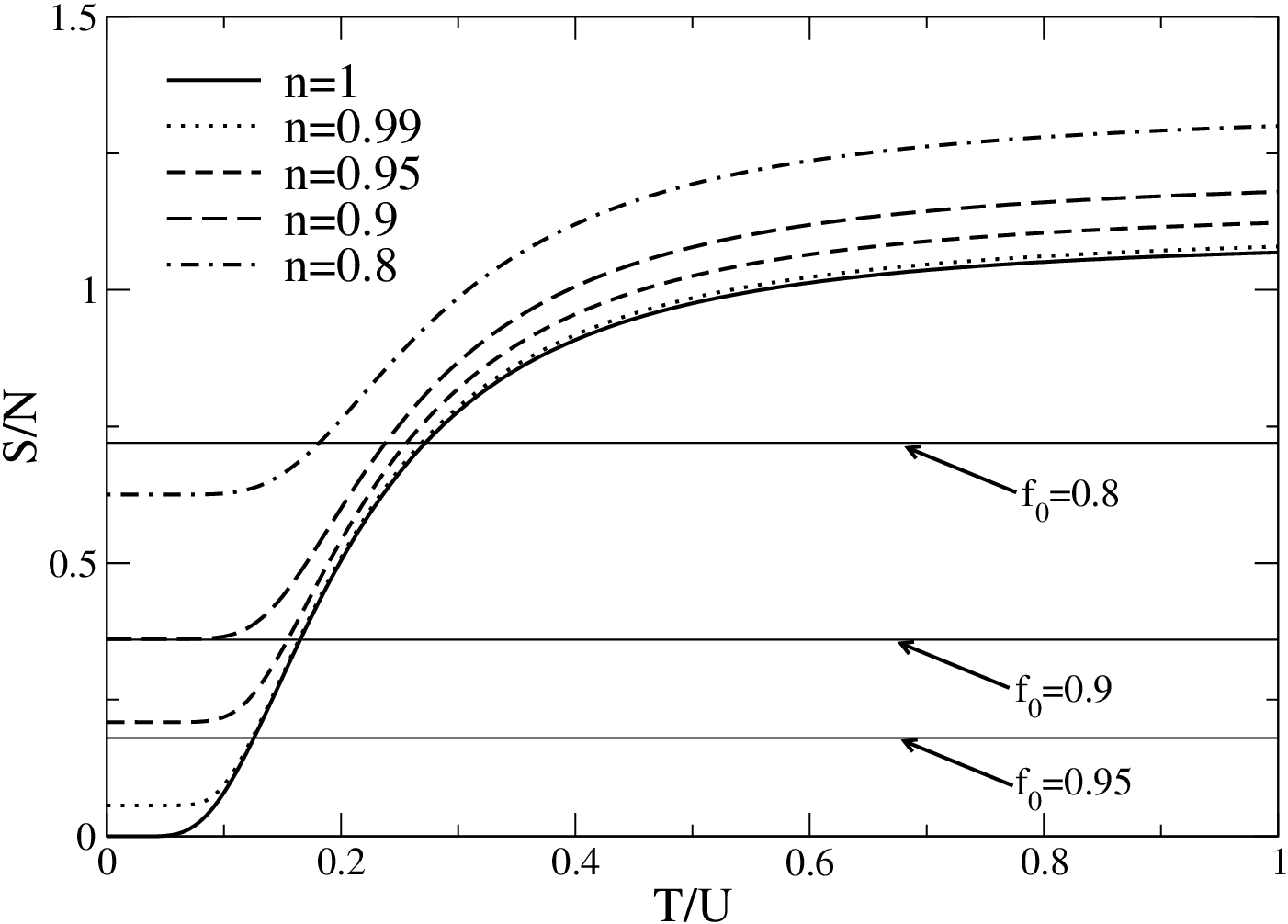}
  \end{center}
  \caption{Entropy per particle as a function of temperature in the atomic 
    limit $t/U=0$ for various values of doping. The entropy per particle of 
    the three-dimensional harmonic oscillator for condensate fractions 
    $f_0=0.95;0.9;0.8$ is included as black horizontal lines.}
  \label{fig:atomiclimite:sgegenT}
\end{figure}

The Gibbs energy of a single site is given by
\begin{equation}
 g = - T \ln\left[ 1+e^{\beta\mu}+e^{-\beta (U-2\mu)}
 \right]\ ,
\end{equation}  
where the terms in the argument of the logarithm refer to the empty, the
singly occupied and the doubly occupied site. 
Introducing the doping $\delta:=n-1$ and
using $\left.\partial_\mu g\right|_T=-n=-(1+\delta)$ 
we arrive at the following expression for the
chemical potential at given doping $\delta$
\begin{equation}
 \mu = \frac{1}{\beta}\ln\left[
 \frac{\delta e^{\beta U }}{2(1-\delta
  )}\left(1-\sqrt{1+ \left(4/\delta^2-4\right) e^{-\beta U }} \right)\right] .
\end{equation} 
The entropy per particle is calculated as the derivative of the
Gibbs energy with respect to temperature $S/N=S/(nL)=
-\left.\partial_T f\right|_\mu$.  In Fig.\ \ref{fig:atomiclimite:sgegenT} the 
entropy is shown for $n\in\{1;0.99;0.95;0.9;0.8\}$. Additionally, the entropy 
for free bosons with a superfluid fraction $f_0=\{0.95;0.9;0.8\}$ according to 
Eq.\ \ref{eq:app:en:SproN} is given. All curves
display an almost constant plateau at small temperatures $T/U<0.1$, an
uprise at about $T/U\sim 0.3$ and a saturation for large temperatures. 

Supposing that the bosons are loaded adiabatically, 
three different regimes can  be identified for fixed $f_0$. In the following, 
the experimentally relevant value $f_0=0.9$ is discussed. At zero doping, the 
temperature is approximately $U/6$, i.e., of the same order as deduced by 
analysing recent spectroscopic experiments \cite{reisc05,stofe04}. 
This seemingly large value
for the temperature is needed because the groundstate of the Mott insulating
phase at zero doping is separated by a large gap from the first excited states.
The entropy has to be produced by the excited states which have an 
energy of the order $U$. 

A small amount of doping leaves this temperature almost unchanged. But the 
system develops a finite entropy at zero (and small) temperatures. The reason 
for this is the degeneracy of the  groundstate coming from the different 
possibilities to distribute $N$ bosons  over $L$ lattice sites. The entropy at 
intermediate values of the temperature is  only changed slightly. The 
temperature realized in the optical lattice is  changed abruptly when the zero 
temperature entropy reaches the entropy 
value of the free bosons. Then the temperature immediately jumps to zero. 
Physically this corresponds to the situation where the whole entropy can be 
produced by the groundstate degeneracy of the doped Mott insulator. Thereby 
the temperature is reduced to the scale $t$ which is zero in the atomic
limit. This is the mechanism which is most probably observed in the experiment 
in Ref.\ \onlinecite{pared04}. It is more closely examined in the next part. 

In the  third temperature regime, the entropy of the free bosons is smaller 
than the entropy of the adiabatically loaded bosons at any temperature. 
This means that an adiabatic loading of the bosons into the
optical lattice is not possible. Instead, a non-ergodic mixture will
be realized. We emphasize, however, that this feature results from
the macroscopic ground-state degeneracy in the atomic limit which
represents an extreme and idealized situation.

\subsection{Tonks-Girardeau Regime}
\label{ssec:fermions}

The experiment performed in Ref.~\onlinecite{pared04} probes the so-called 
Tonks-Girardeau (TG)  regime \cite{jorda28,girar60,tonks36,lieb63}. 
This is the limit  of large $U$. Only empty and singly occupied sites are 
present and the problem  can be mapped in one  dimension onto fermions without 
interaction  if we consider nearest-neighbor hopping only. If we relax the 
condition of nearest-neighbor hopping the mapping to fermions can still be done
but the resulting system is not without interaction. Hence we focus here on 
nearest-neighbor hopping so that we have a benchmark system which is simple to 
analyze.

In order to know which  temperature is realized  after adiabatic loading in the
TG regime one needs to  know the entropy of the  non-interacting fermion gas.
In one dimension the dispersion of tight-binding fermions with hopping $-t$ 
reads 
\begin{equation}
  \epsilon(k)=-2t \cos(k).
\end{equation}
\begin{figure}[t]
  \begin{center}
    \includegraphics[width=\columnwidth]{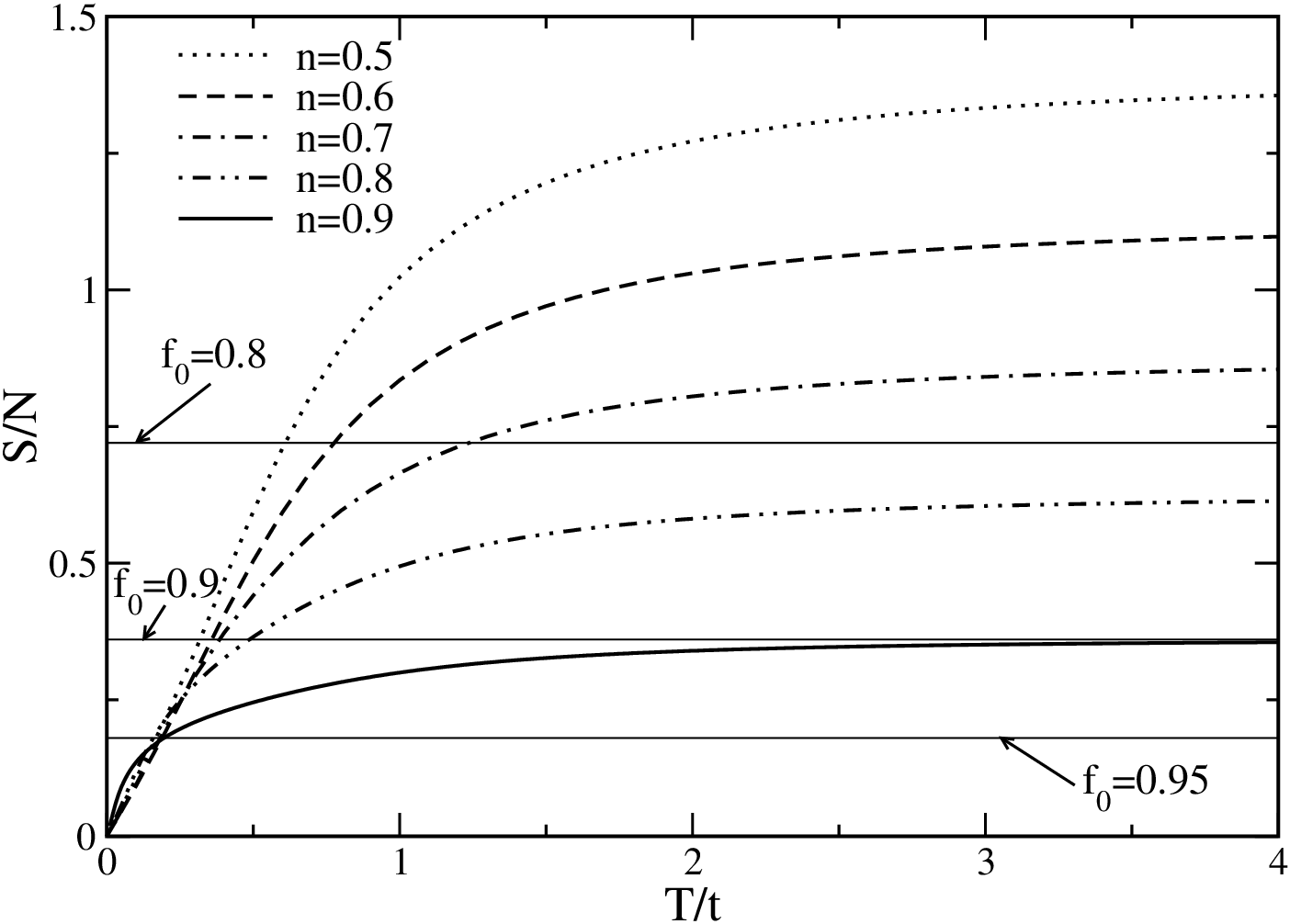}
  \end{center}
  \caption{Entropy of non-interacting tight binding fermions corresponding to
    the TG regime for various fillings. The entropy per 
    particle of the three-dimensional harmonic oscillator for condensate 
    fraction $f_0=0.95;0.9;0.8$ is included as black horizontal lines.}
    \label{fig:tonks}
\end{figure}
The Gibbs energy per site is ($k_{\rm B} =1$)
\begin{eqnarray}
  g&=&\frac{G}{L}=-\frac{ T}{L} \sum_k \ln(1+e^{-\beta \epsilon(k)})
  \\
  &=& -\frac{T}{2\pi} \int_{-\pi}^{\pi} \ln(1+e^{-\beta \epsilon(k)}) dk
\end{eqnarray}
where $L$ is the number of sites. The entropy per particle is calculated as 
\begin{eqnarray}
  \frac{S}{N}&=&\frac{1}{n}\frac{S}{L}=-\frac{1}{n} \left.\frac{\partial
    g}{\partial T} \right|_\mu
  \\
  \label{appendix:bose:ffentropy}
  &=&\frac{k_b}{\pi}\int_{-\pi}^{\pi}
  \left(
  \ln(1+e^{-\beta \epsilon(k)})+\frac{\epsilon(k)}{T(1+e^{-\beta \epsilon(k)})}
  \right)dk\quad .\nonumber
\end{eqnarray}
In the limit of high temperatures the entropy per site is 
\begin{equation}
  \label{appendix:bose:highT}
  {S}/{L}= -n  \log (n )-(1-n ) \log (1-n )
\end{equation}
which is just the entropy of one state with occupation probability $n$. 
The entropy per particle is accordingly
\begin{equation}
  \label{appendix:bose:highT2}
  {S}/{N}=- \log (n )+(1-{n}^{-1}) \log (1-n ).
\end{equation}
Figure \ref{fig:tonks} shows the entropy per
particle versus $T/t$ in the TG regime for fillings $n \in 
\{0.5,0.6,0.7,0.8,0.9\}$.  After a steep increase the entropy reaches its 
saturation value of $- \log (n ) +(1-{n}^{-1})\log (1-n )$. Filling 
$n=0.5$ shows the largest saturation value. 
If the filling is increased further the saturation value decreases. 
Note that the $S/L$ is particle-hole symmetric, i.e., it does not matter whether we consider the filling of fermions $n$ or the filling of holes $1-n$.

Additionally, the entropy per particle of the three-dimensional harmonic trap
is plotted for condensate fractions $f_0=0.95;0.9;0.8$ as horizontal lines. 
This allows to determine the temperature that results for these condensate
fractions after loading into the optical lattice in the TG regime.
For half filling $n=0.5$, the temperatures are $T/t=0.17;0.32;0.61$ for
$f_0=0.95;0.9;0.8$. Also temperatures of multiples of the hopping parameter
are possible, e.g., for $f_0=0.838$ and $n=0.8$ the temperature is $T/t=2$.  
Again, the temperatures are of the size of the microscopic system parameter 
$t$.  This agrees well with the temperatures estimated in Ref.\
\onlinecite{pared04}. There, the temperature has been determined from the
measured momentum profiles.\

For higher filling the entropy saturates at a lower value. 
For filling $n=0.8$ it does not rise up to the value of the harmonic trap
for $f_0=0.8$. Thus in a process of adiabatic loading one cannot achieve a
filling of $n=0.8$ when starting with a condensate fraction of $f_0=0.8$.
Within the TG regime, even the highest temperatures are not sufficient
to reach the necessary entropy. In practice this means that a system starting
from too high an initial entropy will not end up in the TG regime.
Doubly occupied sites will occur corresponding to temperatures of the
order of the corresponding gap. This means that the description as
non-interacting gas of fermions will no longer hold.

\section{Continuous Unitary Transformation}
\label{sec:cut}

After having discussed the atomic limit and the case of free fermions, we study
the Bose-Hubbard Hamiltonian at finite $t/U$ by means of
a particle conserving CUT at zero temperature \cite{reisc05}. We reformulate 
the model in terms of new quasi-particles above the reference state with one 
boson per site. These particles obey hardcore statistics. Aspects of the method
will be presented in the next subsection. The focus is again laid on 
thermodynamic, mainly entropy, considerations. We adapt a formalism introduced 
for spin ladders  \cite{troye94} to approximate the hardcore statistics to 
obtain results at finite  temperature \cite{reisc05}. In subsection 
\ref{ssec:finiteT} 
this  formalism is introduced for the case of the Bose-Hubbard Hamiltonian. In 
particular, we check the reliability of the approximate hardcore statistics by
comparison to the results of non-interacting fermions in the TG regime. At 
the end of this part, we present results for the entropy and the temperature of
the undoped and doped Mott insulating phase for finite $t/U$.

\subsection{Method}
\label{ssec:method}

In this part some technical aspects concerning the CUT are presented. For the
general method we refer the interested reader to Refs.\
\onlinecite{knett00a,knett03a,reisc04,reisc05}. To set up the CUT calculation 
the reference state has to be fixed.  For $t/U \to 0$  and  filling $n=1$ the 
groundstate of Eq.~\ref{eq:bose:hamiltonian} is the product  state of precisely
one boson per site
\begin{equation}
  \label{eq:bose:ref}
  |\text{ref}\rangle=|1\rangle_1\otimes |1\rangle_2 \ldots\otimes |1\rangle_N,
\end{equation}
where $|n\rangle_i$ denotes the local state at site $i$ with $n$ bosons. We
take $|\text{ref}\rangle$ as the reference state. For finite $t/U$ the kinetic
part $t H_{t}$ will cause fluctuations around the
reference state. Deviations from $|\text{ref}\rangle$ are considered to be
elementary excitations. Each site can be occupied by an infinite number of
bosons. The local Hilbert space has infinite dimensions. Therefore, there
are also infinitely many linear independent local operators on this space. To 
set up a real space CUT, it is convenient to truncate
the otherwise proliferating number of terms \cite{reisc04,reisc05}.  
In the present work, the local bosonic Hilbert space is
truncated to four states. Numerical studies using density-matrix
renormalization group have shown that this does not change
the relevant physics \cite{pai96,kashu96,kuhne98}. The CUT gives reliable
results in a wide parameter regime up to $t/U\approx 0.2$, see Ref.\ 
\onlinecite{reisc05}.  Since the calculation is restricted to four local states
including the reference state there are three local creation operators. Applied
to the reference state $|\text{ref}\rangle_i=|1\rangle_i$ they create local
excitations. A hole at site $i$ is created by 
$h^\dagger_i |1\rangle_i= |0\rangle_i$, a particle by 
$p^\dagger_i |1\rangle_i= |2\rangle_i$, and 
$d^\dagger_i |1\rangle_i= |3\rangle_i$ induces another kind of 
particle at site $i$. The particle created by $d^\dagger_i$ corresponds to a
local occupation of a site by three bosons which is treated as an independent
quasi-particle. The operators $h$, $p$ and $d$ obey the 
commutation relations for hardcore bosons. The local bosonic operator
$b^\dagger_i$ is thus represented by the $4\times 4$ matrix
\begin{equation}
  \label{eq:boson:bdagger}
  b^\dagger_i=h^{\phantom \dagger}_i+\sqrt{2}p^\dagger_i+\sqrt{3}
 d^\dagger_i p^{\phantom\dagger}_i
\to \left(
    \begin{array}{cccc}
      0 &0&0&0\\
      1 &0&0&0\\
      0 &\sqrt{2}&0&0\\
      0 &0&\sqrt{3}&0\\
    \end{array}
\right)_\text{\large \it i}
\ ,
\end{equation}
which acts on the vector of the coefficients of 
$\{ |0\rangle,|1\rangle,|2\rangle,|3\rangle\}$.
The CUT is implemented using this presentation of the Hamiltonian with finite 
local matrices \cite{reisc04}. 
%
%
The Hamiltonian Eq.~\ref{eq:bose:hamiltonian} conserves the number of 
bosons. But if it is rewritten in terms of $h,p, d$ it does not conserve the
number of these newly introduced particles. For example, the application of
$H_{t}$ to $|\text{ref}\rangle$ generates a particle-hole pair. To eliminate
the terms that change the eigenvalue of $H_{U}$ an infinitesimal generator 
of the CUT is chosen as  
\begin{equation}\label{boson:eta}
  \eta_{i,j} (l)={\rm sgn}\left(q_i-q_j\right)H_{i,j}(l)
\end{equation}
in an eigenbasis of $H_{U}$; $q_i$ is the corresponding eigenvalue of
$H_{U}$, see Ref.~\onlinecite{knett00a}. The choice of this generator leads
to an effective Hamiltonian commuting with $H_{U}$ which implies that the
processes changing the number of quasiparticles are eliminated.
For the bosonic Hubbard model a complete elimination cannot be achieved
due to the occurrence of life time effects \cite{reisc05}. But this effect
is of minor importance in the parameter regime under study.

The bottom line is that the CUT maps the complicated elementary excitations
of the bosonic Mott insulator onto hardcore particles. The corresponding
annihilation (creation) operators are $p^{(\dagger)}$, $h^{(\dagger)}$,
and $d^{(\dagger)}$.

\subsection{Finite Temperatures}
\label{ssec:finiteT}

In order to discuss thermodynamic properties of the Bose-Hubbard model we have
to extend the calculation to finite temperatures. The CUT yields an effective
Hamiltonian. The particles involved obey hardcore commutation
relations. The statistics of hardcore particles, however, is complicated and
thus a direct evaluation of the partition function is not easily tractable. In
order to obtain results at finite temperature we will adapt the 
calculation in Ref.\ \onlinecite{troye94} for hardcore triplet excitations,
 called triplons, to our present purposes. 

The explicit derivation of the approximate Gibbs energy for hardcore bosons is
presented in Appendix~\ref{sec:app}. The resulting Gibbs
energy per particle reads 
\begin{equation}
\label{eq:bose:freeenergyImText}
g=-(T/L)\ln \tilde{Z}=- T \ln(1+z^p+z^h),
\end{equation}
where we use  partition functions for the individual particles 
defined in Eq.\ (\ref{eq:bose:kleinepartfu}).

\subsection{Test of the Approximate Thermodynamics}
\label{ssec:test}

We check the  validity of the approximate thermodynamics. 
The results for the TG regime from the
previous section, Sect.~\ref{ssec:fermions}, are used as a benchmark. 
In particular, the quality of the approximate statistics
for low temperatures and fillings $n\neq 1$ is addressed.\\
\begin{figure}[t]
  \begin{center}
    \includegraphics[width=\columnwidth]{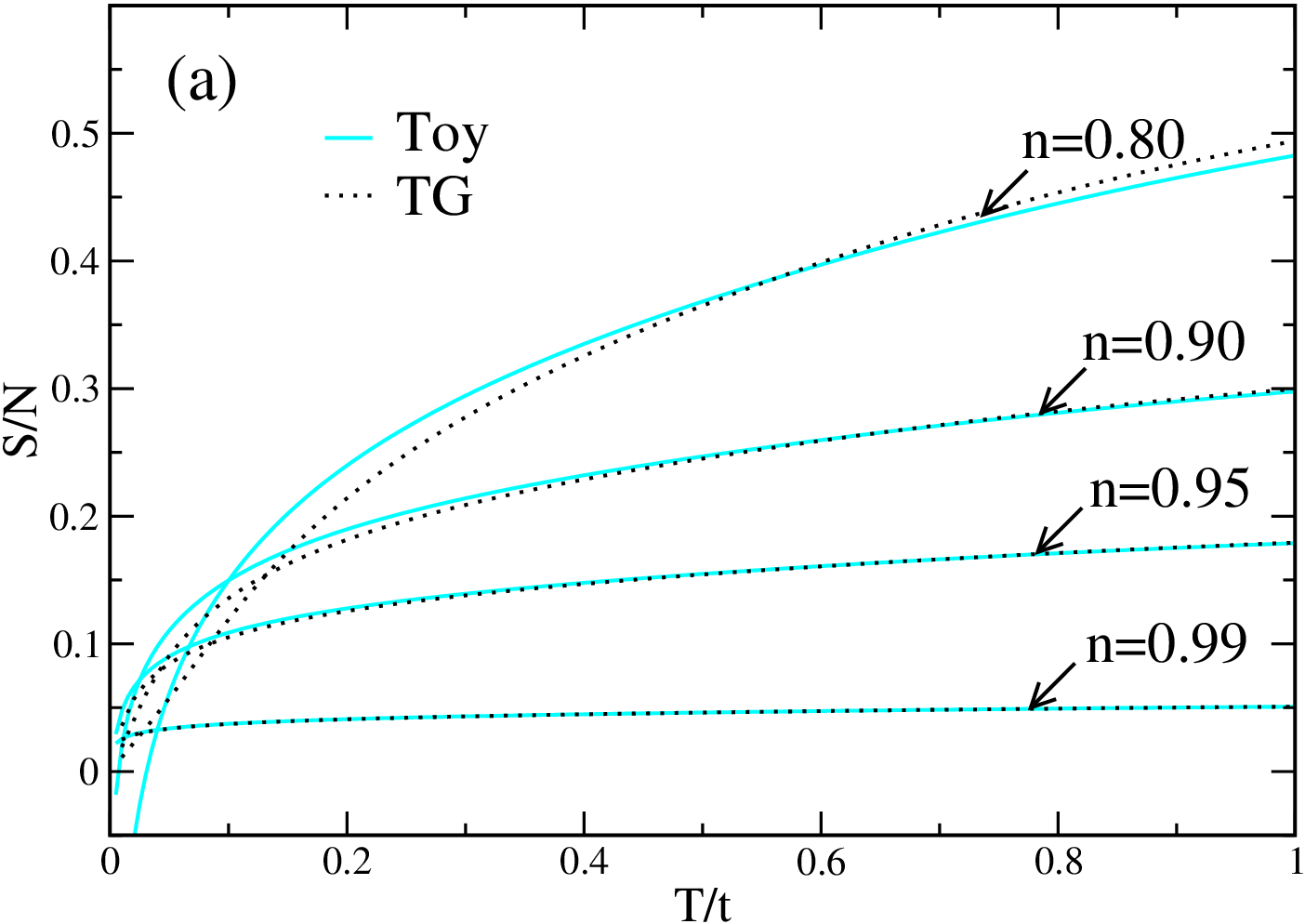}
    \includegraphics[width=\columnwidth]{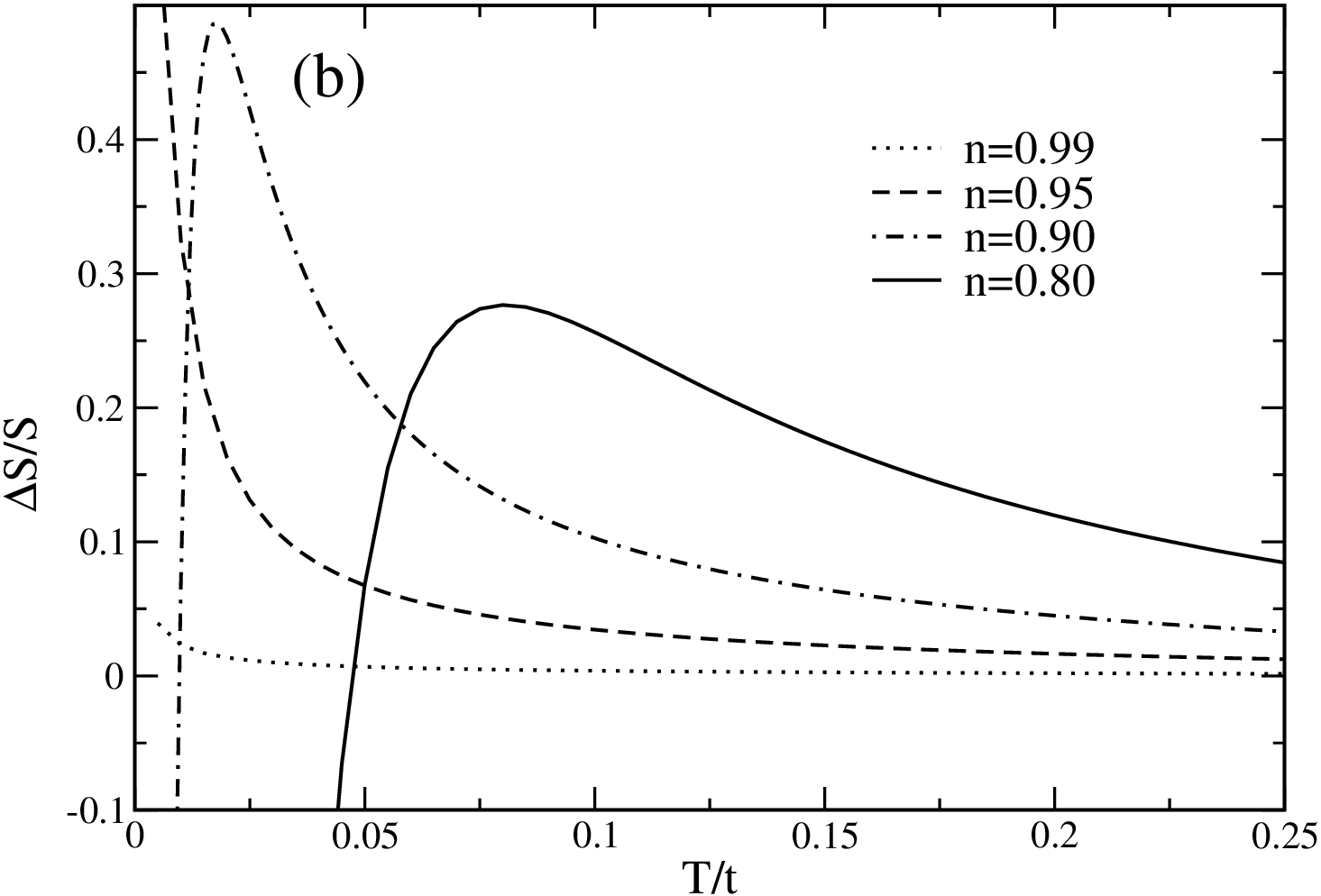}
  \end{center}
  \caption{Comparison of entropies per particle at small temperatures 
    calculated for the TG regime. (a) (color online) 
    Entropy of free fermions is shown as dotted black curves. Results from 
    the toy model calculation in gray (cyan). For fillings close to $n=1$ the 
    entropies show good agreement. But deviations increase with increasing
    doping and for decreasing temperature.
    (b) Relative deviation $\Delta_S=(S_{\text{Toy}}-S_{\text{TG}})
    /S_{\text{TG}}$ of the entropy of the toy-model from the TG result.}
    \label{fig:boson:smallT}
\end{figure}
For this purpose, we introduce a toy model which corresponds to
the TG regime. The CUT calculation deals with the situation found for the 
bosonic Hubbard model for filling $n=1$ and for values $t/U$ in the 
Mott insulating phase. Above the groundstate of singly occupied sites there
exist particle- and hole-like excitations. The toy model includes only the 
hole-like particles. Its Gibbs energy in Eq.~\ref{eq:bose:freeenergyImText} 
simplifies accordingly (cf.\ Appendix A)
\begin{equation}
\label{eq:bose:freeenergyToy}
g_{\text{toy}}=- T \ln(1+z^h)\ .
\end{equation}
 The dispersion contains only a single cosine-term
\begin{equation}\label{eq:bose:dispToy}
\omega^{h}(k)= -2t \cos(k)
\end{equation}
corresponding to hardcore bosons with nearest-neighbor hopping only. With 
these definitions the toy model describes the TG regime. 

\begin{figure}[t]
    \begin{center}
     \includegraphics[width=\columnwidth]{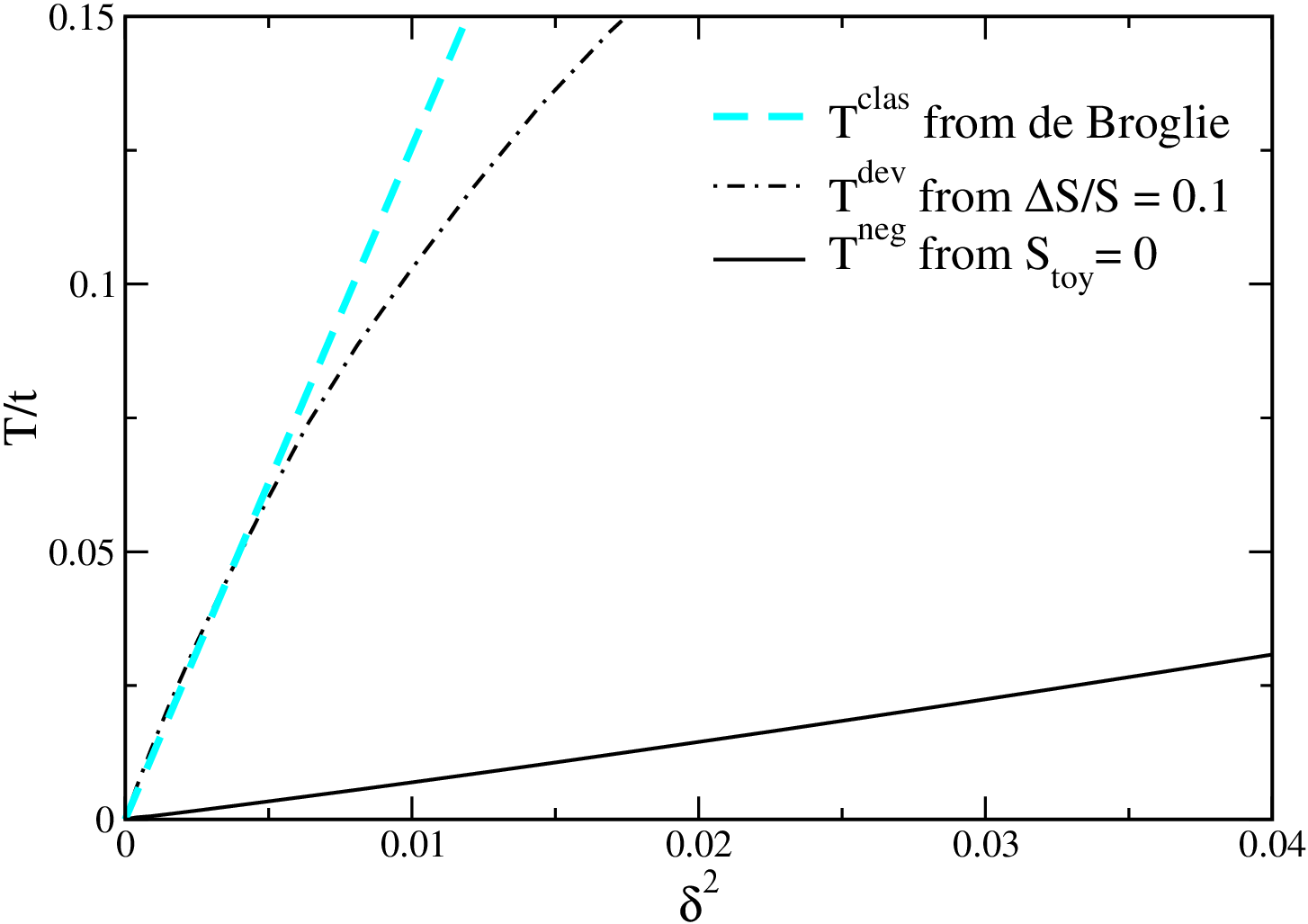}
   \end{center}
    \caption{(color online) The toy model calculation differs from
     the exact fermionic calculation at low temperatures. The
     temperatures $T^{\rm dev}$ and $T^{\rm min}$ characterizing the limit of 
     the approximate thermodynamics are plotted versus $\delta^2=(1-n)^2$.
     For comparison, $T^{\rm clas}$ resulting from the thermal de Broglie
     wave length $\lambda$ is included. For further details see main text.
     }
    \label{fig:boson:T_abweichung}
\end{figure}

The results for this toy model calculation are depicted in Fig.\
\ref{fig:boson:smallT}a in comparison to the entropy from the exact formula for
non-interacting fermions. Discrepancies are found to be increasing with 
doping. They appear especially at low temperatures.
 
Since the deviations are small the  relative deviation $\Delta S/S=
(S_{\text{toy}}-S_{\text{TG}})/S_{\text{TG}}$ is  plotted in Fig.\ 
\ref{fig:boson:smallT}b. For $n=0.8$ the difference  of the entropies has a 
maximum at $T/t\approx 0.08$, then it drops and becomes even negative at very 
low temperatures because $S_{\text{toy}}$ becomes negative for $T\to 0$.
This is an unphysical artifact of the approximate thermodynamics for hardcore 
bosons. 
For $n\rightarrow 1$ the maximum of the relative deviation is lower and shifted
to lower temperatures. But artificial negative entropy behavior is still found 
for very low $T$. 

To investigate systematically, below which temperature the approximate 
thermodynamics becomes significantly wrong, we define the temperature 
$T^{\rm dev}$, at which the relative deviation $\Delta S/S$ exceeds $10\%$. It 
is displayed in Fig.\ \ref{fig:boson:T_abweichung} as a function 
of  $\delta^2=(1-n)^2$. In addition, the temperature $T^\text{neg}$ is shown 
which is defined as the  temperature where the entropy of the toy model becomes
negative.  The temperatures  $T^\text{dev}$ and $T^\text{neg}$ characterize the
limit of applicability of the approximate  thermodynamics.

Both temperatures are compared in Fig.\ \ref{fig:boson:T_abweichung} to the 
temperature $T^\text{clas}$ which results from equating the thermal de Broglie
wave length 
\begin{equation}
\lambda_T = \hbar \sqrt{\frac{2\pi}{m k_\text{B}T}}
\end{equation}
with the average distance $1/\delta$ between two hole excitations $h$.
The inverse mass $1/m$ is given by $2t$ which is the value of the 
second derivative at the band minimum. The excellent agreement in the
proportionality $T\propto \delta^2$ and the very good agreement of 
$T^\text{clas}$ and $T^\text{dev}$ clearly show that the approximate
thermodynamics is a classical treatment, which is valid as long as the gas of 
excitations is dilute enough.
This argument quantifies the limitations of the approximate thermodynamics. 
For very low temperatures the results become unphysical. 
On the other hand, it shows that the results obtained for intermediate
and high temperatures agree very well with the exact behavior in the TG regime.

\subsection{Results}
\label{ssec:results}
 
In Sect.~\ref{sec:limits}, we have already learned that the temperature of the 
bosons in the optical lattice depends strongly on the regime of the system
(Mott insulator or TG regime). The crucial point is whether the energetically
low-lying states realize the initially
present entropy or not. In the following, we  investigate the entropy
at finite $t/U$ for the effective model obtained by the CUT
calculation and evaluated at finite temperatures by the
approximate statistics for hardcore bosons.

In the limit of small $t/U$, the CUT can be performed straightforwardly. 
The results for the Mott phase are obtained for commensurate filling $n=1$.
For fillings $n\neq 1$ the reference state used for normal ordering
before truncation should ideally be the disordered state at the lower filling
$n$ \cite{reisc05,knett03a}. 
This requires to redo the whole CUT for each $n$. But the effect of a
different reference state is not very large if $t/U$ is small. 
So our approach is to perform the CUT with the reference state 
(\ref{eq:bose:ref}) at $n=1$ as described in detail in 
Ref.\ \onlinecite{reisc05}. 
Note that this approach is exact as long as the CUT is performed without
any approximation. So it does not represent an additional approximation in 
itself. The deviation from $n=1$ is included only on the 
level of the approximate thermodynamics of hardcore bosons described in
Sect.~\ref{ssec:finiteT}. By tuning the chemical potential in Eq.\
\ref{eq:bose:kleinepartfu} the average number of holes can be changed.
This way of accounting for doping works particularly well for $n\approx 1$.

\begin{figure}[t]
    \begin{center}
     \includegraphics[width=\columnwidth]{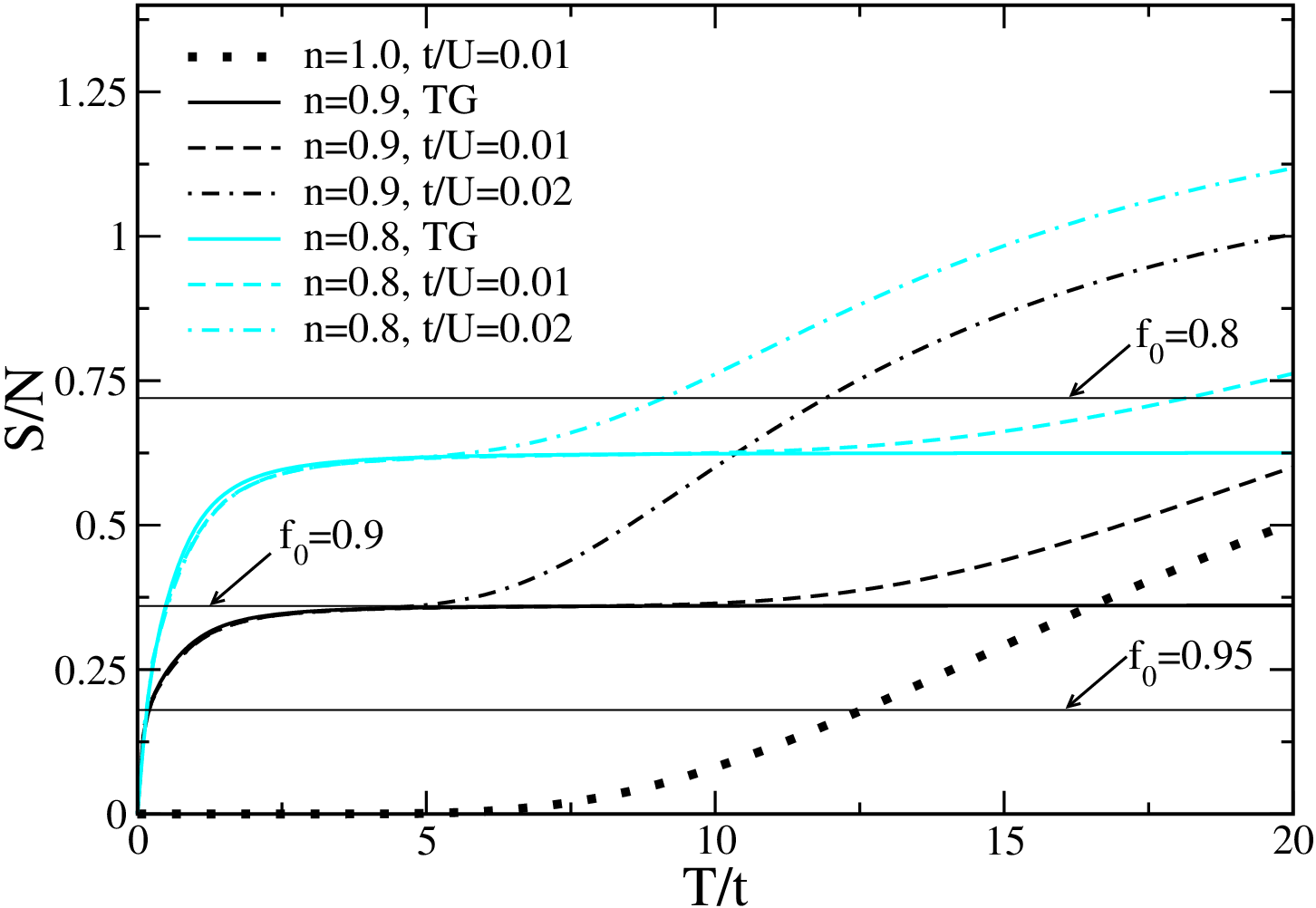}
   \end{center}
    \caption{(color online) Entropy per particle in the TG regime. For small 
      temperatures, the results from the CUT and the free Fermion calculation 
      agree. At higher temperatures there are deviations caused by the presence
      of double occupancies in the bosonic Hubbard model.}
    \label{fig:boson:scompare}
\end{figure}

The entropy per site for the Bose-Hubbard system is calculated from the Gibbs 
energy (\ref{eq:bose:freeenergyImText}) as 
$s=S/N=-\left. \frac{\partial g}{\partial T}\right|_\mu$.
First, the entropy in the TG regime is discussed. In Fig.\ 
\ref{fig:boson:scompare}, the results for the entropy
per particle from the CUT are compared to the results for the TG regime. 
The figure shows results for $t/U=0.01$ and $t/U=0.02$ corresponding to
large values of the 
parameter $\gamma = U/t$ where the TG gas is realized \cite{pared04,wesse05}. 
The thick dotted curve shows the entropy per particle for $t/U=0.01$ for
filling $n=1$. The increase of entropy with temperature is much slower for
$n=1$ than for fillings $n=0.9$ (black curves) and $n=0.8$ (gray/cyan
curves). For filling $n=0.9$, a very good agreement is found for low 
temperatures. The results based on CUT and on the approximate thermodynamics
reproduce nicely the entropy in the TG regime. 
Good agreement is found up to temperatures of
$T/t\approx 10$. For these temperatures, the TG result has already
saturated at its high temperature limit. On the other hand, the CUT results
begin to feel the presence of particle states (double occupancies in terms
of the original bosons) which are not included in the TG description. 
Thus the CUT results deviate from the TG ones. The entropy is 
plotted versus $T/t$ in Fig.\ \ref{fig:boson:scompare}. A temperature of $10t$
corresponds to $T/U=0.1$ for $t/U=0.01$ but to $T/U=0.2$ for $t/U=0.02$. This
is why the results for $t/U=0.02$ show deviations already for  $T/t\approx
5$. A similar behavior is seen for $n=0.8$. 

The small deviations discernible at low temperatures ($T\approx t$) must be 
attributed to the difference between the effective model obtained from the CUT
and the simple fermionic tight-binding model with nearest-neigbor hopping only
in the TG regime.

\begin{figure}[t]
    \begin{center}
     \includegraphics[width=\columnwidth]{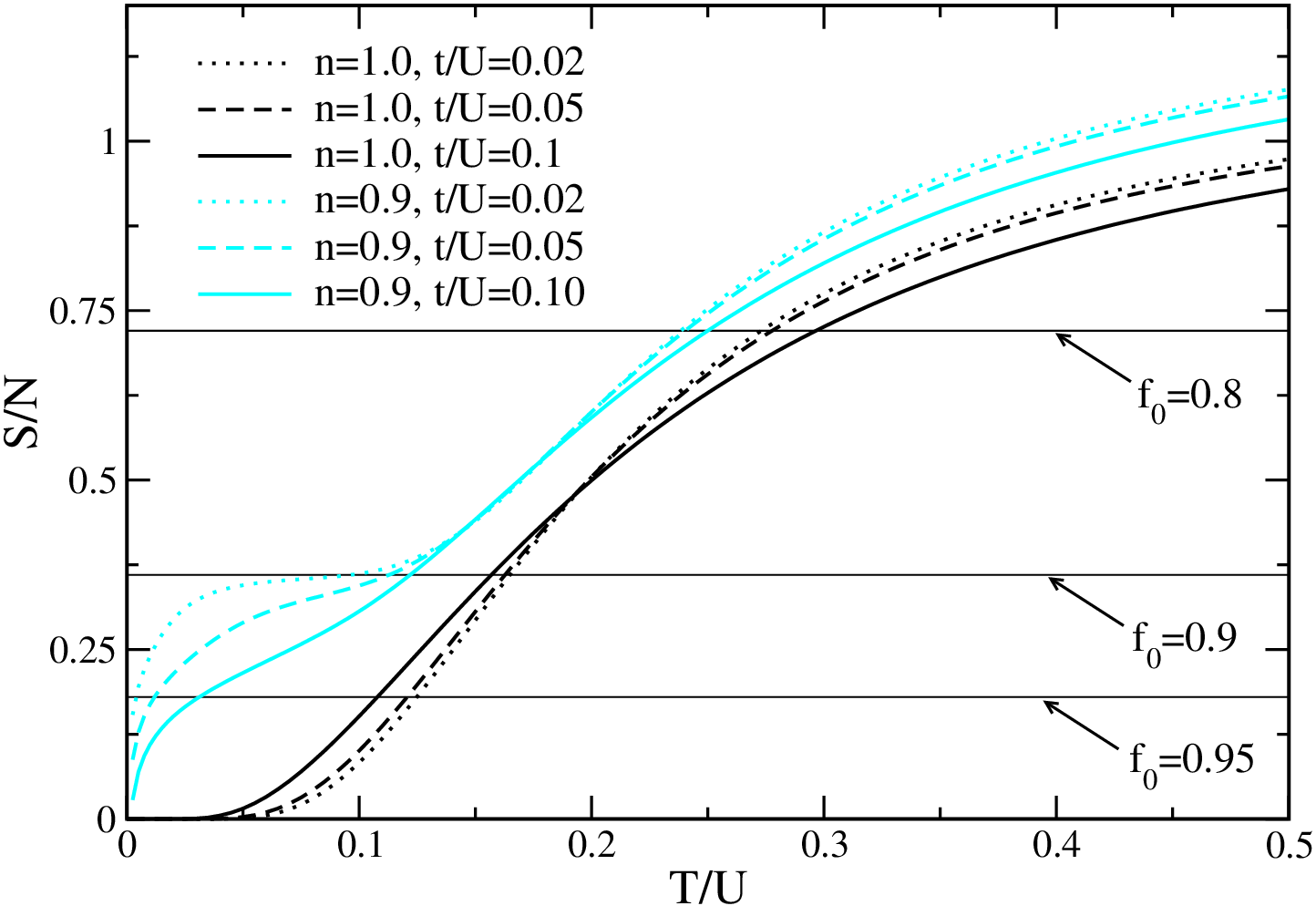}
   \end{center}
    \caption{(color online) 
      Entropy per particle as function of temperature for 
      $t/U={0.02,0.05,0.1}$ at zero doping (black curves) and finite doping
      $n=0.9$ (cyan/gray curves). In addition, the value for the entropy per 
      particle of the three dimensional harmonic oscillator with condensate 
      fraction $f_0={0.8,0.9,0.95}$ is given. }
    \label{fig:boson:sgegenT}
\end{figure}

Finally, Fig.\ \ref{fig:boson:sgegenT} displays the entropy from the CUT 
calculation as function of  temperature in units of $U$.  The entropy per 
site for the undoped case is shown in Fig.\ \ref{fig:boson:sgegenT} for three 
values of $t/U$ (black curves). The entropy does not change significantly for 
the values of $t/U$ that are shown because all of them are still deep in the 
Mott insulating phase.

The entropy per particle for condensate fractions $f_0=\{0.95;0.9;0.8\}$ is 
included as horizontal lines in Fig.~\ref{fig:boson:sgegenT}. Assuming such 
condensate fractions and a hopping parameter $t/U=0.02$ the corresponding
temperatures are 
$0.12;0.16;0.27$ in units of the interaction parameter $U$. These values are
within the range of temperatures that were found by the analysis of spectral
weights \cite{reisc05}. 

The influence of doping is shown in Fig.~\ref{fig:boson:sgegenT} by cyan
(gray) curves for $n=0.9$. Clearly, several differences can be
identified. At finite doping and small $t/U$, the constant entropy
plateau seen for $t/U=0$ in Fig.\ \ref{fig:atomiclimite:sgegenT} can still be 
recognized but there are changes for small temperatures of the order $t$. There
is a finite dispersion at any finite $t$. Therefore, the entropy at zero
temperature is zero because the macroscopic groundstate
degeneracy is lifted. For increasing $t$ the plateau is washed out. At very low
temperatures, there is a sharp uprise from zero entropy to a reduced entropy
value comparable to the value realized at the plateau in the atomic limit.
At higher temperatures, the entropy is dominated by the particle states 
(double occupancies) in analogy to what was found in the undoped case.

\section{Conclusions}
\label{sec:conclusion}

The main purpose of this work was to clarify the different orders of magnitude
of  temperature in recent experiments on the Mott insulating phase of ultracold
 bosons loaded adiabatically in one-dimensional optical lattices. On the one 
hand, temperatures of the order $U$ are found for the undoped Mott insulating 
phase \cite{stofe04,reisc05}. On the other hand, temperatures of the order 
$t$ are found in the TG regime.

By thermodynamic considerations, we found a physically relative simple and 
consistent picture. In the undoped Mott insulator, the temperature has to be
of the  order $U$. The groundstate is a single state and does not provide any 
entropy.  The initially present entropy has to be realized by excited 
states. This requires a large temperature of the order of the gap. The 
situation is different for the doped Mott insulator. Already at the energy 
scale of the hopping an appreciable amount of entropy is possible due to the 
finite amount of holes or particles.

In comparison to experiment, an item still missing in our treatment is the 
inhomogeneity of the trap. So one might wonder whether our qualitative 
conclusions apply to the realistic systems. This question can be answered
positively because it has been shown previously that the 
groundstate displays constant density plateaus at small values of 
$t/U$ \cite{batro02,kolla04}. For zero or low temperature,
 the realistic situation can be seen as an ensemble of homogeneous systems of 
various fillings  \cite{batro05}. Thus it is expected that our qualitative
conclusions concerning the orders of magnitude of temperature remain valid
even if the inhomogeneity of a wide trap is taken into account.

The results were obtained on the one hand by considering limiting cases where 
a purely analytic treatment is possible and on the other hand by using a 
particle conserving CUT at zero temperature extended by an approximate
thermodynamics  for hardcore bosons. For the CUT the Bose-Hubbard 
Hamiltonian is reformulated in terms of a reference state with exactly one 
boson per site and excitations above this state. This approach yields reliable 
results in a wide parameter regime of the Mott insulating phase. Results at 
finite doping were obtained  by tuning the chemical potential in the 
approximate hardcore thermodynamics. The reference state was not adapted to 
finite doping, which may represent a way to improve the CUT calculation. 
But we do not expect significant changes.

We  tested the reliability of the approximate thermodynamics for
hardcore bosons by comparison to the exact results in the TG regime which are
based on the mapping of the system to non-interacting fermions.
We find that the approximation works well up to
relatively low temperatures $T>t/2$. We found that the approximate 
hardcore statistics corresponds to a classical description of the
gas of hardcore particles. This results from the observation that the
temperature below which the approximate statistics fails can be estimated
to be the temperature at which the thermal de Broglie wave length $\lambda_T$ 
equals the interparticle distance. 

Our considerations were restricted to one dimension because only in 
one dimension the hardcore bosons can be exactly mapped to fermions
providing an easy-to-test benchmark. But apart from this generic
one-dimensional feature used for the benchmark
none of our arguments is restricted to one dimension. 
Thus we expect that the qualitative findings presented
in this paper apply also to higher dimensions.

The obtained results provide a consistent picture of the role of temperature 
for bosons in one-dimensional optical lattices. We found that different 
physical situations can produce rather different temperatures. It is 
important in future experimental and theoretical works to take the influence
of temperature into account in order to explore new quantum phenomena in the
field of ultracold atoms.

\section{Outlook}\label{sec:outlook}

Our basic finding is that a restricted phase space at low energies
leads inevitably to a relatively large temperature when the system is
reached adiabatically from a Bose-Einstein condensate with a
condensate fraction which deviates from unity. One obvious means to reach
lower temperatures is to realize condensates with less entropy, i.e.,
with a higher condensate fraction $f_0$ at lower temperature. 

Let us, however, suppose that this route cannot be used to reach
the desired very low temperatures which are, for instance, necessary
to perform reliable quantum computations. In such a case, we like to propose
an alternative route. The system with very restricted phase space
at low energies should be brought into weak thermal contact with another system
in the optical trap which has a large phase space at low energies.
Elementary thermodynamics tells us that the temperatures of both systems 
will finally become equal. This means that the system with large phase
space cools the wanted system with small phase space down to low temperatures.

In the realm of bosons loaded in optical lattices this proposal means that
one system should be realized which corresponds to a Mott insulator and
which remains at relatively high temperatures in the adiabatic process
because it does not have phase space below its gap.
This Mott insulator can be cooled by a system which is in the TG
regime, i.e., in the doped Mott insulating regime, with small hopping 
amplitude $t$. The TG system can accommodate for an appreciable
amount of entropy even at low temperatures. Thus it is well-suited to cool
the other system.

We cannot devise an experimental system which realizes the above idea.
The use of two different laser systems in spatial vicinity for the same kind
of atoms is certainly cumbersome. So it might be more promising to
use two kinds of atoms in the same optical trap. The control parameters
for the first kind of atoms should be such that it becomes
Mott insulating upon loading into the optical lattice. The second kind should
enter the TG regime upon loading. It is most challenging
to see to how low temperatures a bosonic Mott insulating system can be 
pushed in such a way.

\begin{acknowledgments}
Fruitful discussions are acknowledged with A.\ L\"auchli, T.\ St\"oferle, I.\ 
Bloch, S.\ Dusuel, and E.\ M\"uller-Hartmann. 
This work was supported by the DFG in SFB 608 and in SP 1073.
\end{acknowledgments}

\appendix

\section{Approximate Thermodynamics for Hardcore Bosons}\label{sec:app}

This appendix provides the derivation of the approximate Gibbs energy of 
hardcore particles used in Sect.\ \ref{sec:cut}. The calculation adapts the 
reasoning in Ref.\ \onlinecite{troye94} for triplets in a spin ladder to
the present hardcore excitations. The relevant excitations are independent 
particle and independent hole excitations. Multiple correlated $p$ 
and $h$ excitations and
excitations of $d$-type are less relevant because of their higher energy. 
Therefore, we will neglect $d$-excitations and correlated excitations of more 
than one particle or hole in the following.  

To use the 
non-interacting boson statistics for hardcore particles is a bad approximation 
because non-interacting bosons have no restriction on the occupation number. 
An incorrect number of states  would contribute in the partition sum. 
The idea is to reweight all contributions to the partition sum such that
they contribute the correct amount of entropy.  Therefore we have to calculate 
the correct number of states with particle and hole excitations. 

The number of possibilities to distribute $M$ truly bosonic excitations on 
$2L$ sites is    
\begin{equation}
  \label{eq:boson:plainboson}
  g_\text{B}(L,M)= \binom{2L+M-1}{M}.
\end{equation}
But the correct number of possibilities of $M$ hardcore excitations of two 
kinds reads
\begin{equation}
  \label{eq:boson:hcb}
  g(L,M)=2^M \binom{L}{M}.
\end{equation}
To obtain the correct entropy for each number of excitations the $M$-boson part
in the partition sum is reweighted by the factor 
$g(L,M)/g_\text{B}(L,M)$. By this 
means, the partition function displays the correct high and low temperature 
behavior.  The boson partition function for a system with $L$ sites and $M$ 
bosons is
\begin{equation}
  \label{eq:boson:smallpf}
  Z^\prime(L,M)=\!\!\!\! \sum_{\stackrel{\{k_j,k_l\}}{M=M_p+M_h}}\!\!\!\! e^{ 
    -\beta \left(\sum_{j=1}^{M_p} \omega^p(k_j)-\mu^p
    +\sum_{l=1}^{M_h}   \omega^h(k_l)-\mu^h\right)}  .
\end{equation}
The sum runs over all multi-indices $\{k_j,k_l\}$ of $M=M_p+M_h$ momenta for
particles at the momenta $\{k_j\}$ and holes at the momenta $\{k_l\}$. 
Here, $\omega^p(k_j)$ is the energy of a 
particle with momentum $k_j$, $\omega^h(k_l)$ the energy of a hole with 
momentum $k_l$. 
The chemical potential $\mu$ is accounted for by $\mu^p=\mu$ and $\mu^h=-\mu$.
The negative sign in the definition of the chemical potential $\mu^h$
for holes results from the fact that adding a hole means removing an original
boson.
The boson partition function Eq.~\ref{eq:boson:smallpf} 
contains $(2L)^M$ terms. Thus the reweighted partition function reads
\begin{subequations}
  \label{eq:boson:pf}
  \begin{eqnarray}
    \tilde{Z}&=&\sum_{M=0}^{L}  \frac{g(L,M)}{(2L)^M} Z^\prime(L,M)\\
    \nonumber
    &=& \sum_{M=0}^{L}  L^{-M} \binom{L}{M}
    \!\!\!\! \sum_{\stackrel{\{k_j,k_l\}}{M=M_p+M_h}} \\
    && \qquad e^{ 
      -\beta \left(\sum_{j=1}^{M_p} \omega^p(k_j)-\mu^p
      +\sum_{l=1}^{M_h}   \omega^h(k_l)-\mu^h\right)} \quad
    \\
    &=& \left[ 1+L^{-1} \sum_{k;\sigma\in\{p,h\}} 
      e^{-\beta(\omega^\sigma(k)-\mu^\sigma)}\right]^L
    \\
    &=& \left[ 1+ z^p + z^h\right]^L\ ,
    \label{eq:boson:pf:final}
  \end{eqnarray}
\end{subequations}
where we used the definition
\begin{equation}
  \label{eq:bose:kleinepartfu}
  z^\sigma := \frac{1}{2\pi}\int_0 ^{2\pi} 
  e^{-\beta(\omega^{\sigma}(k)-\mu^\sigma)}dk
\end{equation}
for the partition function of the single modes with $\sigma\in\{p,h\}$. 
The chemical potential $\mu$ is
 determined by specifying a certain filling.  For instance, for 
the filling $n=1$ 
the chemical potential $\mu$  is determined such that as many particles as 
holes are excited.

We emphasize that Eq.\ \ref{eq:boson:pf} is only an approximation to the 
correct hardcore partition function because it contains contributions from
states that are only allowed for non-interacting bosons, but are forbidden 
for hardcore bosons. The correct  contribution of each subspace
of fixed total number $M$ of excitations is ensured only in an overall
manner by reweighting the total subspace. 

Using the final expression (\ref{eq:boson:pf:final}) 
one can compute the Gibbs energy per site $g$ to be  
\begin{equation}
  \label{eq:bose:freeenergy}
  g=-(T/L)\ln \tilde{Z}=- T \ln(1+z^p+z^h).
\end{equation}
Equations~\ref{eq:bose:kleinepartfu} and \ref{eq:bose:freeenergy} are the
formulae used in Sect.\ \ref{ssec:finiteT}.

The toy model discussed in Sect.\ \ref{ssec:test} includes only holes 
as possible excitations. Thus the Gibbs energy depends only on one partition
function of a single mode $  g_\text{toy}=- T \ln(1+z^h)$.
The dispersion of the toy model contains one single cosine-term
$  \omega^{h}(k)= -2t \cos(k) $
which makes the analytical calculation of $z^h$ possible
\begin{equation}
z^h = e^{\beta \mu^h} I_0(2\beta t)\ ,
\end{equation}
where $I_n(z)$ is the modified Bessel function of the first kind. 
The entropy per particle reads  
\begin{eqnarray}\label{eq:bose:entropyToy}
  \frac{S}{N} &=& -\frac{1}{n}\left.
  \frac{\partial g}{\partial T}\right|_\mu\\
  &=&   -\beta e^{\beta \mu^h}
  \frac{2 t I_1\left(2\beta t\right) + \mu^h I_0\left(2\beta t\right)}
       {n\left(1+e^{\beta \mu^h} I_0\left(2\beta t\right)\right)} \nonumber \\
       && \qquad
       + \frac{1}{n}\log\left(1+e^{\beta \mu^h} I_0\left(2\beta t\right)
       \right)\ .\quad
\end{eqnarray}


\end{document}